\documentclass{ws-procs9x6}
\usepackage{amsmath}
\usepackage{amsfonts}
\usepackage{amssymb}

\begin{document}
\title{Strong Decays of $N$ and $\Delta$ Resonances in the Point-Form Formalism
}
\author{T. MELDE, W. PLESSAS, and R.F. WAGENBRUNN
}
\address{Institut f\"ur Physik, Theoretische Physik\\
Universit\"at Graz, \\ 
Universit\"atsplatz 5,
A-8010 Graz, Austria\\ 
E-mail: thomas.melde@uni-graz}
\maketitle
\abstracts{
We present covariant predictions of relativistic constituent quark models
for $\pi$ and $\eta$ decay widths of $N$ and $\Delta$ resonances. The 
results are calculated for a model decay operator within the 
point-form spectator approximation. It is found that most theoretical
values underestimate the experimental data considerably. 
}
\section{Introduction}
It has long been of interest to describe strong decays
of baryon resonances with realistic constituent quark models (CQM).
Most of the modern calculations have investigated the effects of the
dynamics of the underlying CQM, the nature of the decay operator
(e.g., elementary emission vs. quark pair creation) or relativistic 
corrections~\cite{Capstick:1993th,Geiger:1994kr,%
Stancu:1989iu,Theussl:2000sj}. In addition one has faced considerable
uncertainties regarding the phase-space factor, since the transition
amplitudes have not been derived in a relativistically invariant manner.
Recently, we obtained direct predictions of a relativistic CQM from
a Poincar\'e invariant 
calculation of baryon resonance decays in the pion 
channel~\cite{Melde:2002ga}. The decay operator was based on the 
point-form spectator approximation (PFSA), 
an approach that already proved successful in
the description of the electroweak structure of the 
nucleons~\cite{Wagenbrunn:2000es}.
\section{Theory}
Here, we study the $\pi$ and $\eta$ decays of $N$ and $\Delta$ resonances
in a relativistic approach. Specifically, the theory is formulated along 
Poincar\'e-invariant quantum mechanics~\cite{Dirac:1949,Keister:1991sb}.
We adhere to its point-form version, since it allows to calculate 
the observables in a manifestly covariant manner~\cite{Klink:1998pr}.
This approach, being distinct from a field-theoretic treatment,
relies on a relativistically invariant mass operator with the 
interactions included according to the Bakamjian-Thomas 
construction~\cite{Bakamjian:1953}. There\-by it ful\-fils all 
symmetries required by special relativity.
At this instance, we use a rather simplified model for the 
decay operator, because our first goal is to set up a fully 
relativistic CQM formulation of strong decays of baryon resonances.

Baryons are simultaneous eigenstates of the 
mass operator $\hat M$ and the four-velocity operator $\hat V$
(or equivalently of the four-momentum operator $\tilde P^{\mu}$), the 
total-angular-momentum operator $\hat J$, and its z-component 
$\hat \Sigma$. 
We denote these states by the corresponding eigenvalues
$\left|v,M,J,\Sigma\right>$. 
The transition amplitude for the decays is defined in
a Poincar\'e-invariant fashion, under overall momentum conservation
($Mv_{\rm in}-M'v_{\rm out}=Q_{\pi}$), by
\begin{eqnarray}
F\left({\rm in}\rightarrow {\rm out}\right)
& = &
\left< v_{\rm out},M',J',\Sigma'\right|
       \hat {D}_{\alpha}
       \left|v_{\rm in},M,J,\Sigma\right>
       \nonumber\\
        &\sim & 
\int{      d^{3}k_{2}d^{3}k_{3}d^{3}k'_{2}d^{3}k'_{3}
	     }
	      \Psi^{\star}_{M'J'\Sigma'}
  \left(
  \vec k'_{i}
;\mu'_{i}
  \right)
  \Psi_{MJ\Sigma}
  \left(
  \vec k_{i}
;\mu_{i}
  \right)
	     \nonumber \\
	     & &
	     \prod_{\sigma'_{i}}{
	     D^{\frac{1}{2}\star}_{\sigma'_{i}\mu'_{i}}
	     \left[R_{W}\left(k'_{i},B\left(v_{\rm out}\right)\right)\right]}
	     \prod_{\sigma_{i}}{
	     D^{\frac{1}{2}}_{\sigma_{i}\mu_{i}}
	     \left[R_{W}\left(k_{i},B\left(v_{\rm in}\right)\right)\right]}
	     \nonumber
	     \\
	     & &
	     \left<p'_{1},p'_{2},p'_{3};\sigma'_{1},\sigma'_{2},\sigma'_{3}
	     \right|
	     \hat {D}_{\alpha}
	     \left|p_{1},p_{2},p_{3};\sigma_{1},\sigma_{2},\sigma_{3}
	     \right>,
\end{eqnarray}
where the baryon wave functions $\Psi^{\star}_{M'J'\Sigma'}$ and
$\Psi_{MJ\Sigma}$ enter as representations of the rest-frame baryon states 
$\left<M',J',\Sigma'\right|$ and $\left|M,J,\Sigma\right>$, 
respectively.

In a first attempt,  
we assume a decay operator in the PFSA with a pseudo-vector coupling. 
In particular, the decay operator is given in the form
\begin{eqnarray}
& &
    \left<p'_{1},p'_{2},p'_{3};\sigma'_{1},\sigma'_{2},\sigma'_{3}\right|
    \hat {D}_{\alpha}
    \left|p_{1},p_{2},p_{3};\sigma_{1},\sigma_{2},\sigma_{3}
	\right>
    \nonumber
    \\
& &
    =
    \sqrt{\frac{M^{3}M'^{3}}{
    \left(\sum{\omega_{i}}\right)
    \left(\sum{\omega'_{i}}\right)}}
    3i g_{q\pi}
    \bar u\left({p'_{1}},\sigma'_{1}\right)
    \gamma^{5}\gamma^{\mu}{ \lambda^{\alpha}}
    u\left({ p_{1}},\sigma_{1}\right)
    \nonumber
    \\
& &
    \times
    2{p'}^{0}_{2} \delta\left(\vec p_{2}-\vec p'_{2}\right)
    2{p'}^{0}_{3}\delta\left(\vec p_{3}-\vec p'_{3}\right)
   \delta_{\sigma_{2}\sigma'_{2}}
   \delta_{\sigma_{3}\sigma'_{3}}
   \frac{Q_{\pi\mu}}{2m},
\end{eqnarray}
%
where $g_{q\pi}$ is the pion-quark coupling constant, 
$\lambda^{\alpha}$ 
the flavour operator for the particular decay channel,
$m$ the quark mass, and $M$ as well as $M'$ are the masses of the
decaying resonance and the final nucleon, respectively.
It should be noted that in PFSA the impulse delivered
to the quark that emits the pion is not equal to the impulse
delivered to the baryon as a whole. Thus, the momentum $\tilde q$
transferred to the single quark is a fraction of the momentum 
$Q_\pi$ transferred to the residual nucleon; it is determined uniquely
by the overall momentum conservation and the two spectator conditions.
\section{Results}
In table~\ref{table1} we show the direct predictions for the decay widths
in the pion channel for the Goldstone-Boson-Exchange
(GBE)~\cite{Glozman:1998ag} and 
the One-Gluon Exchange (OGE)~\cite{Theussl:2000sj} CQMs
calculated in PFSA. We also include corresponding results
for the Instanton-Induced CQM (II) obtained in a Bethe-Salpeter
approach~\cite{Metsch:2004qk}. It is clearly seen that all but one 
(the $N_{1535}^*$) decay widths are considerably underestimated
for all of these relativistic models. Furthermore there is a general
trend in the results, namely, the larger the
branching ratio into $\Delta\pi$, the bigger the deviation
of the theoretical results from the experimental data. This becomes 
apparent from the comparison in the last four columns of 
table~\ref{table1}, where we have quoted the $\Delta\pi$ branching 
ratios and expressed the theoretical predictions
as percentage values relative to the experimental widths given by the
PDG~\cite{Hagiwara:2002fs}.

\vspace{-3mm}
\renewcommand{\arraystretch}{1.5}
\begin{table}[hb]
\tbl{
Theoretical predictions for $\pi$ decay widths of various relativistic
CQMs in comparison to experimental data 
\label{table1}
}
{\begin{tabular}{@{}crccccccc@{}}
\hline
\multicolumn{9}{c}{}\\[-2ex]
Decays&Experiment\cite{Hagiwara:2002fs}&\multicolumn{3}{c}{
Rel. CQM
}&
$\Delta\pi$
&\multicolumn{3}{c}{
$\%$ of Exp. Width
}\\
{$\rightarrow \pi N_{939}$}&{}&GBE&OGE&II&{}&GBE&OGE&II\\
[0.25ex]
\hline
\multicolumn{9}{c}{}\\[-2ex]
$N^{\star}_{1440}$
&
$\left(227\pm 18\right)_{-59}^{+70}$ &
$30$ &
$37$ &
$38$ &
$20-30\%$ & 
$13$ &
$16$ &
$17$\\ 
$N^{\star}_{1520}$
&
$\left(66\pm 6\right)_{-\phantom{0}5}^{+\phantom{0}9}$&
$17$ &
$16$ &
$38$ &
$15-25\%$ & 
$26$ &
$25$ &
$58$\\ 
$N^{\star}_{1535}$
&
$ \left(67\pm 15\right)_{-17}^{+28}$&
$93$ &
$123$ &
$33$ &
$<1\%$ & 
$130$ &
$180$ &
$49$\\ 
$N^{\star}_{1650}$
&
$\left(109\pm 26 \right)_{-\phantom{0}3}^{+36}$&
$29$ &
$38$ &
$\phantom{0}3$ &
$1-7\%$ & 
$26$ &
$35$ &
$\phantom{0}3$\\ 
$N^{\star}_{1675}$
&
$ \left(68\pm 8\right)_{-\phantom{0}4}^{+14}$&
$\phantom{0}6$ &
$\phantom{0}6$ &
$\phantom{0}4$ &
$50-60\%$ & 
$\phantom{0}9$ &
$\phantom{0}9$ &
$\phantom{0}6$\\ 
$N^{\star}_{1700}$
&
$ \left(10\pm 5\right)_{-\phantom{0}3}^{+\phantom{0}3}$&
$\phantom{0}1$ &
$\phantom{0}1$ &
$0.1$ &
$>50\%$ & 
$\phantom{0}9$ &
$12$ &
$\phantom{0}1$\\ 
$N^{\star}_{1710}$
&
$\left(15\pm 5\right)_{-\phantom{0}5}^{+30}$&
$\phantom{0}4$ &
$\phantom{0}2$ &
$n/a$ &
$15-40\%$ & 
$27$ &
$15$ &
$n/a$\\ 
$\Delta_{1232}$
&
$\left(119\pm 1 \right)_{-\phantom{0}5}^{+\phantom{0}5}$&
$34$ &
$32$ &
$62$ &
$n/a$ & 
$28$ &
$27$ &
$52$\\ 
$\Delta_{1600}$
&
$\left(61\pm 26\right)_{-10}^{+26}$&
$0.1$ &
$0.5$ &
$n/a$ &
$40-70\%$ & 
$\phantom{0}0$ &
$\phantom{0}1$ &
$n/a$\\ 
$\Delta_{1620}$
&
$\left(38\pm 8\right)_{-\phantom{0}6}^{+\phantom{0}8}$&
$10$ &
$15$ &
$\phantom{0}4$ &
$30-60\%$ & 
$27$ &
$38$ &
$11$\\ 
$\Delta_{1700}$
&
$\left(45\pm 15\right)_{-10}^{+20}$&
$\phantom{0}3$ &
$\phantom{0}3$ &
$\phantom{0}2$ &
$30-60\%$ & 
$\phantom{0}6$ &
$\phantom{0}7$ &
$\phantom{0}4$\\
\hline
\end{tabular}}
\end{table}
\vspace{-3mm}

In table~\ref{table2} we show analogous predictions for the $\eta$ decay 
channel. All but two decay widths are quite
small, what is congruent with the experimental data. However, for the
$N_{1535}^{\star}$ and $N_{1650}^{\star}$ decays the relative
magnitudes are opposite to the ones observed in experiments. In these 
cases there are also appreciable difference between the relativistic
and nonrelativistic results.

In summary the theoretical description of baryon resonance decays is by
no means complete. In addition to the possible improvements expected from a
more refined decay operator, it appears necessary to include explicit 
couplings between different channels. Furthermore, a more realistic 
description of the resonance wave functions beyond pure
three-quark bound states seems to be required.

\vspace{-3mm}
\begin{table}[ht]
\tbl{Theoretical predictions for $\eta$ decay widths of the GBE and OGE
CQMs in comparison to experimental data 
and a nonrelativistic calculation in the elementary emission model (EEM). 
\label{table2}}
{\begin{tabular}{@{}crcccc@{}}
\hline
\multicolumn{6}{c}{}\\[-2ex]
Decays&Experiment\cite{Hagiwara:2002fs}&\multicolumn{2}{c}{
Rel. CQM Models
}
&\multicolumn{2}{c}{EEM GBE
}\\
{$\rightarrow \eta N_{939}$}&{}&GBE&OGE&dir&rec\\
[0.25ex]
\hline
\multicolumn{6}{c}{}\\[-2ex]
$N^{\star}_{1520}$
&
$\left(0.28\pm 0.05\right)_{-0.01}^{+0.03}$&
$0.04$ &
$0.03$ &
$0.03$ &
$0.05$ \\ 
$N^{\star}_{1535}$
&
$ \left(64\pm 19\right)_{-\phantom{0.}28}^{+\phantom{0.}28}$ &
$36$ &
$46$ &
$0.06$ &
$155$ \\ 
$N^{\star}_{1650}$
&
$ \left(10\pm 5 \right)_{-\phantom{0.}\phantom{0}1}
^{+\phantom{0.}\phantom{0}4}$ &
$72$ &
$95$ &
$0.9$ &
$288$ \\ 
$N^{\star}_{1675}$
&
$ \left(0\pm 1.5\right)_{-\phantom{0}0.1}^{+\phantom{0}0.3}$ &
$0.8$ &
$0.8$ &
$0.8$ &
$1.6$ \\ 
$N^{\star}_{1700}$
&
$ \left(0\pm 1\right)_{-\phantom{0}0.5}^{+\phantom{0}0.5}$ &
$0.4$ &
$0.4$ &
$0.2$ &
$0.4$ \\ 
$N^{\star}_{1710}$
&
$\left(6\pm 1\right)_{-\phantom{0.}\phantom{0}4}^{+\phantom{0.}11}$ &
$1.0$ &
$1.4$ &
$0.1$ &
$2.2$ \\
\hline
\end{tabular}}
\end{table}
\vspace{-3mm}
\noindent{This work was supported by the Austrian Science 
Fund (Project P16945).}
%
%

%
%
\end{document}